# A continuous compositional-spread technique based on pulsed-laser deposition and applied to the growth of epitaxial films

H.M. Christen*[,a], S.D. Silliman[†], and K.S. Harshavardhan
*Neocera, Inc., 10000 Virginia Manor Road, Beltsville, MD 20705.*
[*]*Currently at Oak Ridge National Laboratory, Solid State Division, Oak Ridge, TN 37831-6056.*
[†]*Currently at Intel Massachusetts, Inc., Hudson, MA 01749.*

A novel continuous-compositional spread technique based on the non-uniformity of the deposition rate typically observed in Pulsed Laser Deposition (PLD) is introduced. Using rapid (sub-monolayer) sequential deposition of the phase spread's constituents, intermixing of the constituents occurs on the atomic scale during the growth process. Therefore, a pseudo-binary or pseudo-ternary phase diagram is deposited without the requirement of a post-anneal. The approach uses the spatial variations in the deposition-rate naturally occurring in PLD; therefore, there is no need for the masks typically used in combinatorial techniques. Consequently, combinatorial materials synthesis can be carried out under optimized film growth conditions (for example, complex oxides can be grown at high temperature). Additionally, lifting the need for post-annealing renders this method applicable to heat-sensitive materials and substrates (e.g. films of transparent oxides on polymer substrates). PLD-CCS thus offers an interesting alternative to traditional "combi" for situations where the number of constituents is limited, but the process variables are of critical importance. Additionally, the approach benefits from all the advantages of PLD, particularly the flexibility and the possibility to work with targets of relatively small size.
Composition determination across the sample and mapping of physical properties onto the ternary phase diagram is achieved via a simple algorithm using the parameters that describe the deposition-rate profiles. Experimental verification using EDX and RBS measurements demonstrates the excellent agreement between the predicted and the calculated composition values. Results are shown for the high-temperature growth of crystalline perovskites (including $(Ba,Sr)TiO_3$ and the formation of a metastable alloy between $SrRuO_3$ and $SrSnO_3$) and the room-temperature growth of transparent conducting oxides.

## I. INTRODUCTION

Combinatorial chemistry has enjoyed great success in the pharmaceutical research community.[1] Hundreds or thousands of compounds are routinely investigated simultaneously using robotic techniques for synthesis and analysis. In contrast, the application of combinatorial approaches to solid materials in thin film form has been less widespread, largely because the growth conditions of most advanced materials need to be precisely controlled and are not easily reproduced in a combinatorial method. In particular, metal-oxide thin films, such as high-temperature superconductors, ferroelectrics, and magneto-resistive materials, are often processed at elevated temperatures where manipulation of masks is cumbersome and likely to be plagued by contamination problems.

A high-temperature, combinatorial Laser Molecular Beam Epitaxy (LMBE) approach was introduced recently[2] and applied to doping studies of ZnO[3] and the growth of epitaxial superlattices.[4] However, the low deposition rates and the delicate mask alignment procedure currently limit the approach to nine combinations generated per run.

Xiang and co-workers[5] have introduced an approach in which "precursor" films of the desired material's constituents are deposited at room temperature onto a crystalline substrate through a series of masks, then heated to promote inter-diffusion, and finally annealed to crystallize the material. In this approach, the masks are positioned at room temperature, and dozens or hundreds of combinations can be explored in one run. The approach has been successfully applied to magneto-resistive oxides,[6] photoluminescent materials,[7] and ferroelectrics.[8] Each resulting "library" of materials consists of hundreds of individual cells of a known uniform composition, and the observed physical properties can be related to the composition. The effect of several different dopants and the effects of co-doping can thus be observed on one single wafer.

The disadvantage of the approach lies in the fact that the growth method is distinctively different from other techniques usually applied to the growth of these materials.

Continuous compositional-spread (CCS) techniques offer less flexibility in exploring various chemical compositions, because the two-dimensional nature of the film restricts the approach to pseudo-binary and pseudo-ternary systems. However, utilizing composition-gradients naturally occurring in co-evaporation or co-sputtering approaches, CCS can be performed without the need of masks and thus at growth conditions comparable to those that yield optimized materials properties. These approaches were explored as early as the 1960s (see Ref. 9 and references therein), but their success remained limited due to the unavailability of rapid screening and analysis techniques. More recently, a CCS technique based on co-sputtering from three off-axis sources was introduced at Bell Laboratories[10] and was applied to the study of high dielectric-constant materials.[11,12] In this method, growth conditions can closely mimic those encountered in a traditional film growth environment.

[a] Electronic mail: christenhm@ornl.gov

It is clear that a similar approach may be followed using Pulsed Laser Deposition (PLD). Earlier studies have shown that co-ablating from two targets (i.e. using a beam splitter to ablate simultaneously from two sources) can be used to dope $BaTiO_3$ films with cobalt,[13] and an extension of this work to obtain a composition gradient across the substrate would be straightforward. However, if a series of experiments is desired in which each deposited phase spread gradually "zooms in" to a desired composition, the deposition rates for each constituent must be changed by varying the laser energy that impinges on each target. This can easily be accomplished by using apertures, in a way that the laser energy density at the target (which controls the ablation mechanism and has a strong influence on the energetics of the ablated species) remains constant. However, this will result in a change of the spot size, and, consequently, in a change in the spatial distribution of the ablated material. As a consequence, calibration measurements as described below have to be performed each time the range of the deposited phase spread is changed. Alternatively, beam attenuators could be used, having the advantage of preserving the spot shape, but resulting in different energetics for each experiment. Therefore, a parameter of critical importance in PLD has to be changed from one sample to another, possibly introducing erroneous results.

In the present work, a new approach for PLD-based CCS is introduced which suffers from neither of these two handicaps, and "zooming in" onto a desired sub-set of the phase diagram is achieved simply by modifying the number of laser pulses fired onto each material. This method is based on the sequential deposition of sub-monolayer amounts of each constituent, whereby atomic-scale intermixing is achieved instantaneously and under normal PLD growth conditions. Furthermore, adjustment of the composition is achieved by simply varying the ratio of laser pulses fired on each target. For the deposition of uniform-composition thin films, this approach has been applied successfully for materials such as $KTa_xNb_{1-x}O_3$[14] (mixing obtained from sequential ablation of $KTaO_3$ and $KNbO_3$ targets) and $PbTiO_3$[15] (mixing from PbO and $TiO_2$ targets).

This paper first illustrates this new PLD-CCS approach in the case of pseudo-binary phase diagrams, using materials grown at high temperature, which leads to the formation of both conventional and metastable perovskite films. In the second part of this paper, the approach is extended to pseudo-ternary systems, demonstrating the applicability of the method to the room-temperature growth of transparent conducting oxides. The accuracy with which the composition can be predicted for each point of the sample based on simple reference measurements is illustrated.

## II. PSEUDO-BINARY SYSTEMS

The basic approach of PLD-CCS is an automated sequential deposition of sub-monolayer amounts of each constituent, coupled with a mechanical manipulation of the substrate and the target, or of the focusing optics for the excimer laser beam. Figure 1 shows the basic configuration in which the targets and substrate are kept fixed in space, whereas the beam is steered from one target to another: first, a laser pulse (or a fixed number of laser pulses) is fired onto the first target, which is positioned opposite to the substrate but off-center with respect to the middle of the substrate. This results in a deposition of less than a monolayer of the material, with most of the material being deposited nearest to the center of the plume. Repeated firing of the laser in this configuration would result in a film with a non-uniform thickness profile, centered on the axis of the plume.

Immediately after this first sub-monolayer deposition, the laser is focused on a second target, which is positioned again opposite to the substrate, but off-center with respect to the middle of the substrate and positioned to the other side of the substrate. A small number of pulses is then fired onto that second target, and the cycle is repeated.

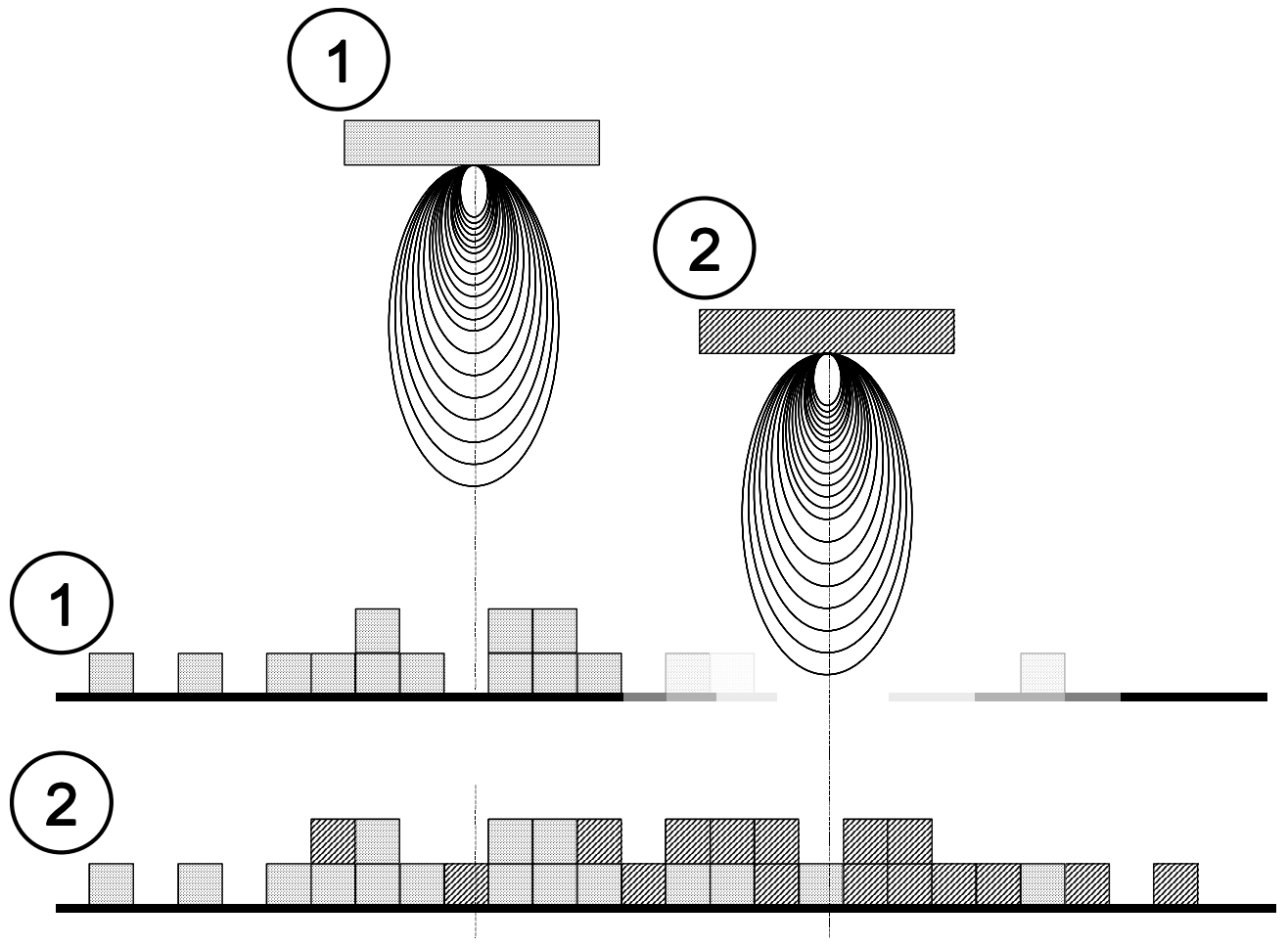

FIG. 1. Schematic representation of the continuous compositional-spread approach using sequential deposition of sub-monolayer amounts of each constituent. Parts of the drawing labeled (1) refer to the first step in the deposition sequence, those labeled (2) to the second step, after which the cycle is repeated until a film of sufficient thickness is obtained.

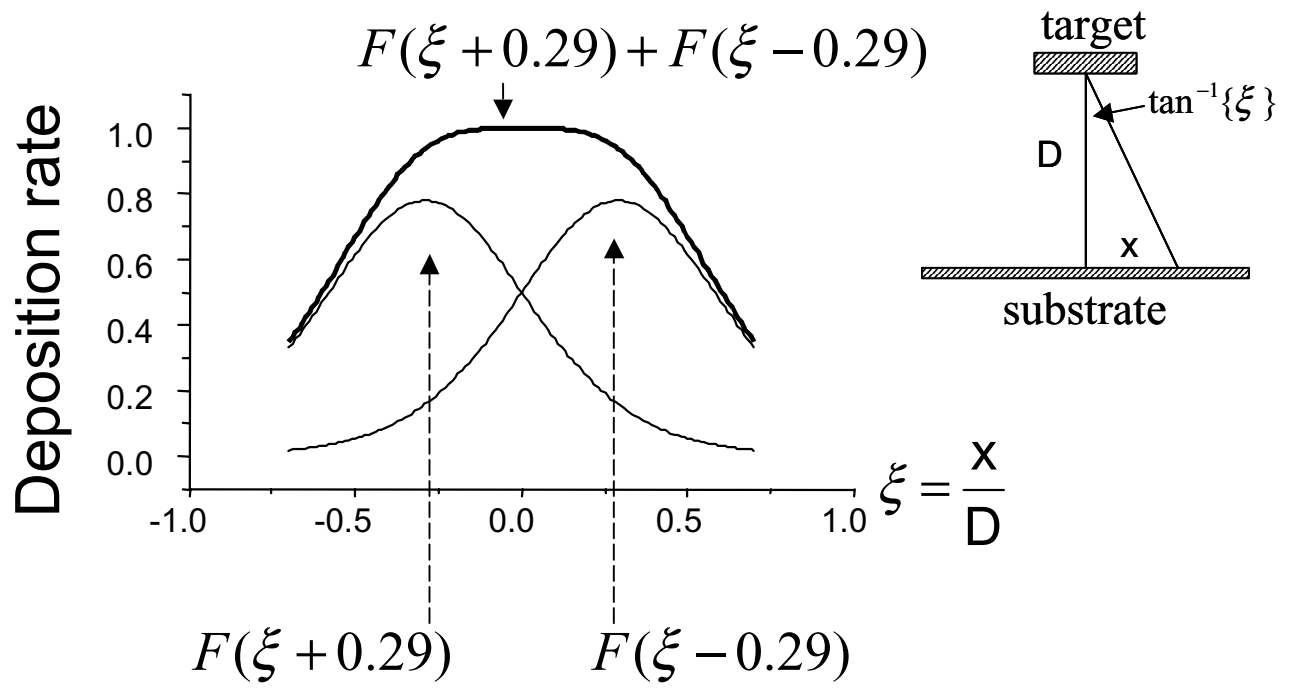

FIG. 2. Calculated deposition rate for a typical angular dependence following a $\cos^n$ behavior. The deposition rate as a function of distance on the substrate is shown on a normalized scale $\xi = x/D$, where $x$ is the position measured on the substrate and $D$ is the target-substrate separation. The sum of two horizontally shifted deposition profiles exhibits a broad flat region in the center of the substrate.

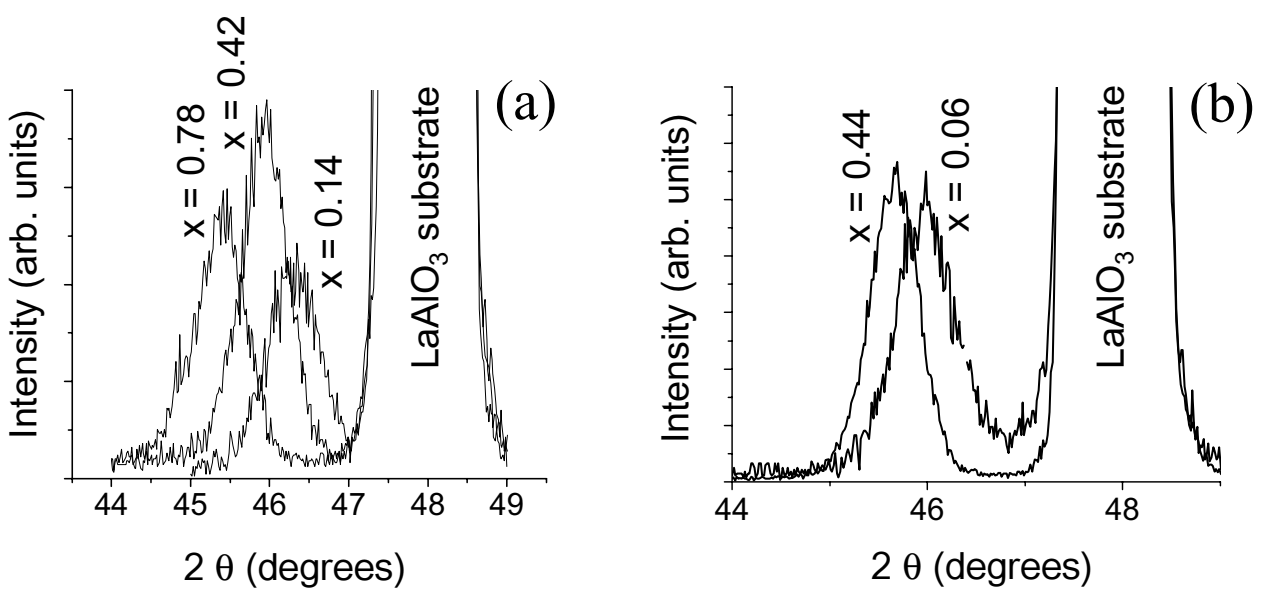

FIG. 3. X-ray θ-2θ scans for epitaxial $Ba_xSr_{1-x}TiO_3$ (a) and $SrRu_xSn_{1-x}O_3$ (b) films obtained by PLD-CCS. Formation of single-phase alloys is achieved by this sequential deposition of sub-monolayer amounts of each constituent for both systems, even though $SrRuO_3$ and $SrSnO_3$ are immiscible in bulk form.

In our experimental configuration, the laser beam was kept stationary, while the substrate was rotated by 180°, and the target was exchanged between each step of the process.

Each step consisting of changing the target, rotating the substrate heater, and firing the laser, requires less than one second. For a typical perovskite material with a unit cell of about 4 Å, and 4 cycles per unit cell, this results in a growth rate of 0.5 Å/s, thus a 100 nm thick film can be grown in just over 30 minutes.

The composition of the material at each point on the substrate can be calculated if the angular dependence of the deposition rate is known. In most PLD experiments, this dependence follows $\cos^n(\theta)$ dependence,[16] where θ is the angle measured from the normal to the target through the laser spot. Therefore, the deposition rate on the substrate at a distance $x$ from the intercept of this normal with the substrate (see inset of Fig. 2) is given by

$$F(\xi,n) = \cos^n(\theta) = \cos^n(\tan^{-1}\{\xi\}) \quad \text{where } \xi = x/D. \qquad (1)$$

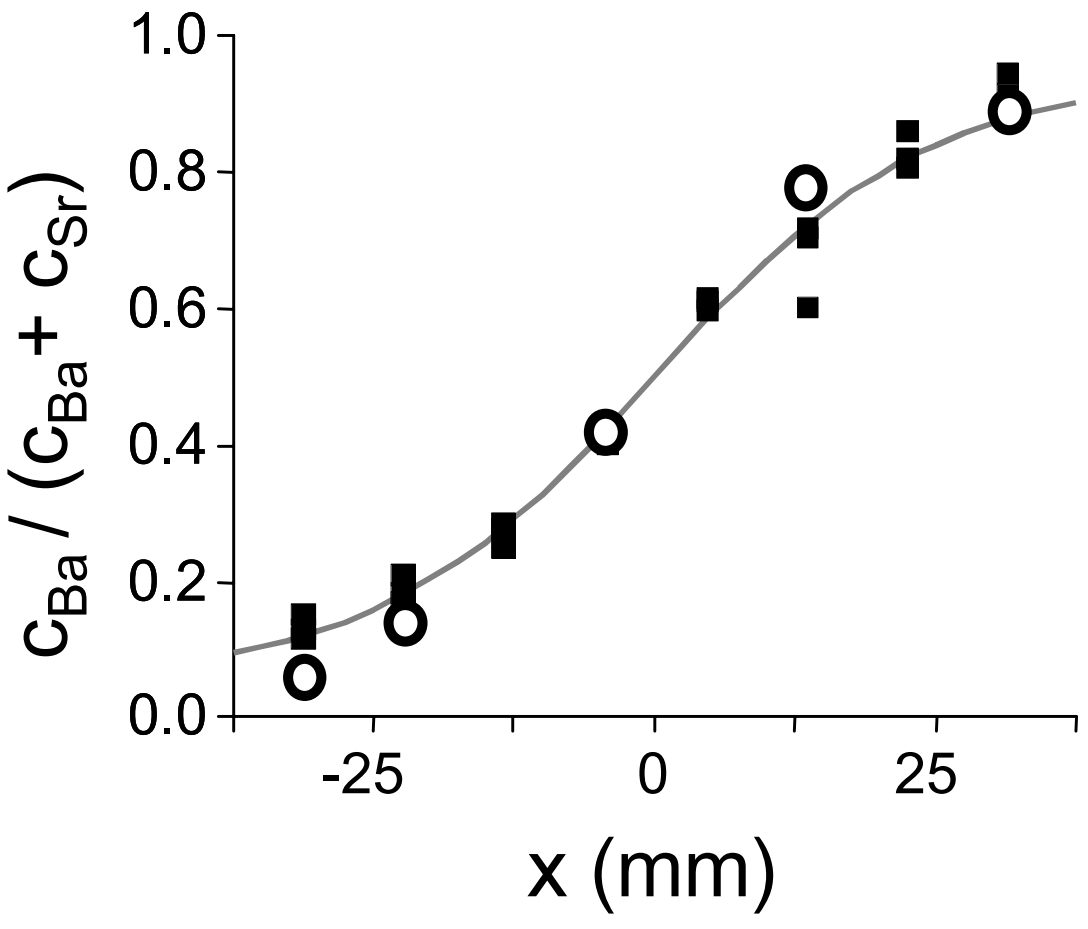

FIG. 4. Composition of $Ba_xSr_{1-x}TiO_3$ films grown by PLD-CCS as a function of position on the heater plate. Open circles are obtained from the x-ray data, solid squares from EDX. The solid line is calculated according to equation (2), using $n$ = 11 as the only free parameter.

$D$ is the distance between the substrate and the target.

Figure 2 illustrates the consequence of this dependence. Two curves, shifted horizontally by an amount $\xi_0$ (i.e. $F(\xi \pm \xi_0, n) = \cos^n[\tan^{-1}\{\xi \pm \xi_0\}]$ ) are drawn together with the resulting total profile. As is clearly seen, for n = 11 and $\xi_0$ = 0.29, the resulting total deposition rate is constant to within less than 5% over a distance equal to half of the substrate-target separation.

The composition at each point of the substrate is easily calculated from this dependence. For example, if material A (exhibiting a deposition rate described by $\cos^{n(A)}$) is deposited with the target positioned at $-\xi_0$ and material B (exhibiting a deposition rate described by $\cos^{n(B)}$) with the target positioned at $+\xi_0$, the composition of the resulting compound $A_{1-y}B_y$ is obtained by calculating

$$y = \frac{c_B}{c_A + c_B} = \frac{F(\xi - \xi_0, n(B))}{F(\xi + \xi_0, n(A)) + F(\xi - \xi_0, n(B))}. \qquad (2)$$

The geometry of the growth apparatus determines $\xi_0$, whereas the exponents n have to be determined from a separate experiment (i.e. the determination of the thickness profile of a film obtained by ablating only from one target).

Figure 3 shows x-ray θ-2θ scans for perovskite films obtained by this method. In each case, eight (001)-oriented $LaAlO_3$ substrates (5mm x 5mm x 0.5mm) were glued along a straight line onto a 3" diameter rotating heater plate using silver paint. The rotation of that heater was synchronized with the exchange of the target and the firing of the laser by a computer.

Figure 3(a) shows the result for $Ba_xSr_{1-x}TiO_3$ grown from $SrTiO_3$ and $BaTiO_3$ targets. The target-substrate distance was 75mm, and the deposition was carried out at 750 °C at 200 mTorr of oxygen. $Ba_xSr_{1-x}TiO_3$ films have previously been studied extensively for applications in DRAMs,[17,18] tunable microwave applications,[19,20] and as high-k gate dielectrics on Si,[21,22] and their properties are well understood. The x-ray data indicates that the films are all (001)-oriented and single-phase, and the composition can be estimated from the peak-positions.

A more direct approach to determining the composition is by using energy-dispersive x-ray spectroscopy (EDX), where films of $SrTiO_3$ and $BaTiO_3$ serve as composition standards. The results of EDX-measurements are shown in Fig. 4 and compared to the numbers obtained from an analysis of the x-ray data. Also shown in Fig. 4 is the prediction from Eq. (2), where the exponent $n$ is the only free parameter (all other parameters are determined by the system's geometry). A very satisfactory agreement is clearly obtained for $n$ = 11. This indicates that for pseudo-binary systems, a determination of the exponent $n$ from a reference measurement is sufficient to determine the composition as a function of position on the substrate heater.

Figure 3(b) shows the x-ray results for films of $SrRu_xSn_{1-x}O_3$, grown from a single-phase $SrRuO_3$ target and a target containing equal amounts of $SrRuO_3$ and $SrSnO_3$.

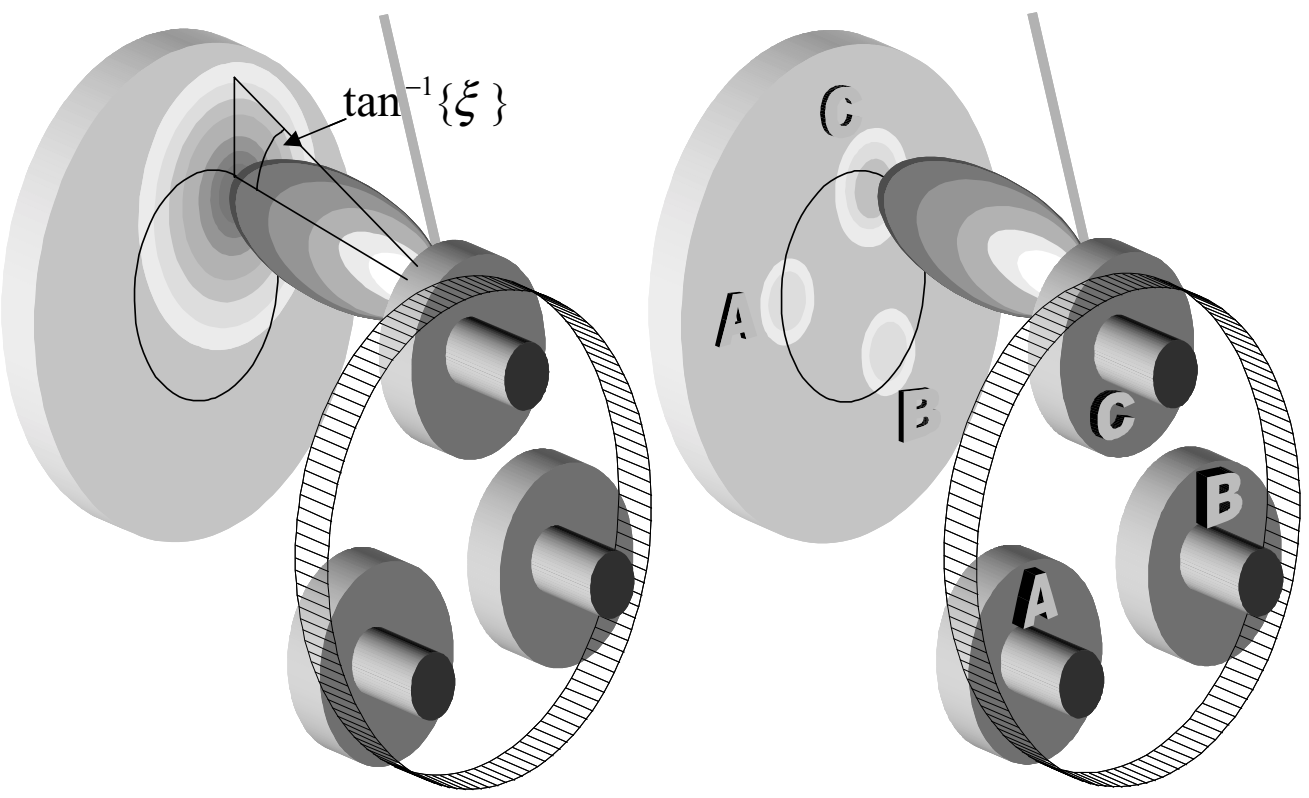

FIG. 5. Schematic representation of the CCS approach for a pseudo-ternary system. A predefined number of laser pulses is fired onto the first target, resulting in a maximum deposition rate at a location away from the center of the substrate. The substrate is then rotated by 120° and the target exchanged. Repeated cycling yields the desired film with varying composition.

Previous studies have shown that these two perovskites are immiscible in the bulk,[23] but that single-phase epitaxial films can be grown from a multi-phase target. The present data shows that this meta-stable alloy can be formed by the sequential deposition of sub-monolayers.

## III. PSEUDO-TERNARY SYSTEMS

The extension of the approach described above to a pseudo-ternary phase diagram is straightforward. Figure 5 shows a schematic drawing of the growth apparatus. Target and substrate are aligned in such a way that the point on the substrate experiencing the highest deposition rate does not coincide with the center of rotation of the heater plate.

In each step of the deposition, a predefined number of laser pulses is fired onto a target, the heater is then rotated by 120°, and the target is exchanged. Repeated cycling yields a film with a composition that varies continuously as a function of position.

The key to success in this PLD-CCS method is the knowledge of the composition at each point on the substrate without having to perform complicated chemical analysis of the sample. To demonstrate that this can indeed be done based on simple thickness measurements on reference samples, a pseudo-ternary phase diagram consisting of $In_{0.95}Sn_{0.05}O_x$ (indium-tin oxide, ITO), ZnO, and yttrium-stabilized $ZrO_2$ (YSZ) was deposited. This system was chosen because all three oxides are readily deposited at room temperature, and Rutherford backscattering (RBS) analysis can easily separate between (In+Sn), Zn, and (Y+Zr).

First, single layers of all of these constituents (ITO, ZnO, and YSZ) were grown on 2” by 3” glass slides. Thickness fringes on these films clearly indicated elliptic thickness contours. The aspect ratio $r = a/b$ of these ellipses was determined by overlapping an optical image of the sample with a drawing of ellipses of known $r$. Profilometry was used to determine the thickness profiles in the direction of the larger axis of the ellipsis, and a least-square fitting to a $\cos^n$ behavior yielded the exponent $n$. Three parameters, namely the exponent $n$, the ellipsis’ aspect ratio $r = a/b$, and the total thickness $T$ at the center of the deposition, completely describe the deposition profile in the entire plane. Figure 6a shows the profile calculated from these three parameters for the case of the ZnO reference sample. Similar profiles were obtained for ITO and YSZ (not shown), rotated by 120° with respect to each other.

From these thickness profiles, the deposition rate for each constituent of the pseudo-ternary system is known, allowing for a determination of the composition and of the total thickness at each point on a mixed film (analogous to Eq. (2)).

The predicted total thickness is shown in Fig. 6b for a case of non-equal amounts of material having been deposited for each of the three constituents. This is achieved by firing a different number of laser pulses onto each target, and results in a sample containing, in this particular case, less ITO than either of the other constituents.

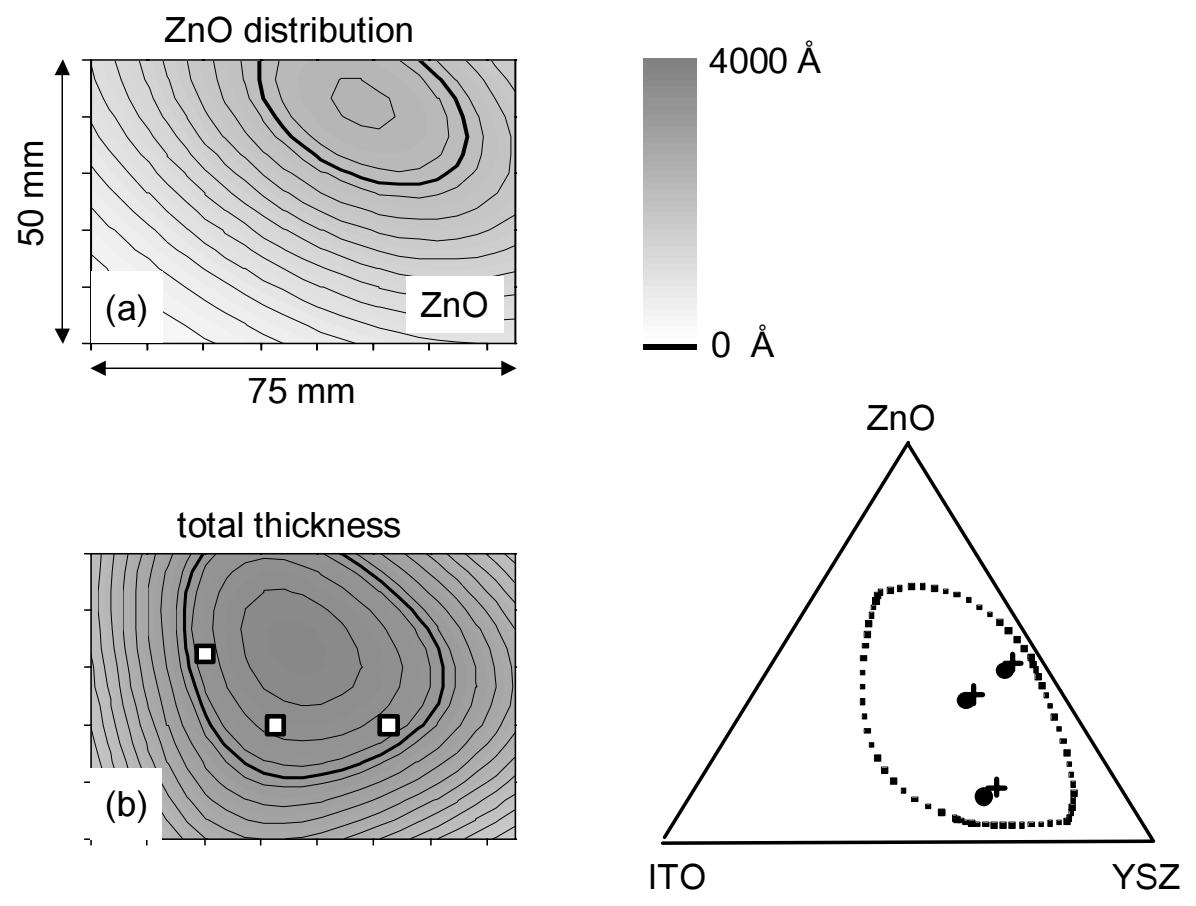

FIG. 6. Calculation of the composition as a function of position on a PLD-CCS pseudo-ternary phase diagram. (a) shows the deposition rate profile across a 2” by 3” substrate calculated from the parameters obtained by fitting to results from profilometry. (b) shows the total thickness resulting from a deposition of three materials (ITO, ZnO, and YSZ) where different amounts are deposited for each constituent (hence the asymmetric shape of the profile). The dotted line on the pseudo-ternary phase diagram indicates the region that is obtained in this single deposition. For the positions marked by squares in (b), the composition was determined by RBS, as indicated by solid circles in the phase diagram. Predictions based on the one-dimensional thickness profiles obtained on reference samples are indicated by crosses, demonstrating the agreement between measurement and calculation.

The composition at each point of the sample is then simply calculated using a computer algorithm, and the dotted line in the phase diagram of Fig. 6 indicates the portion of the phase diagram covered by this single sample. Other portions of the phase diagram, or a more symmetrically positioned fraction of the phase diagrams, can obviously be selected simply by changing the ratio of laser pulses fired onto each target.

Because the $\cos^n$ dependence yields some amount of deposition of each of the constituents onto all parts of the substrate, the end points of the phase diagrams can obviously not be reached by this method. This, however, does not present a significant limitation, as these points are the most easily accessible by simple PLD growth from the respective targets.

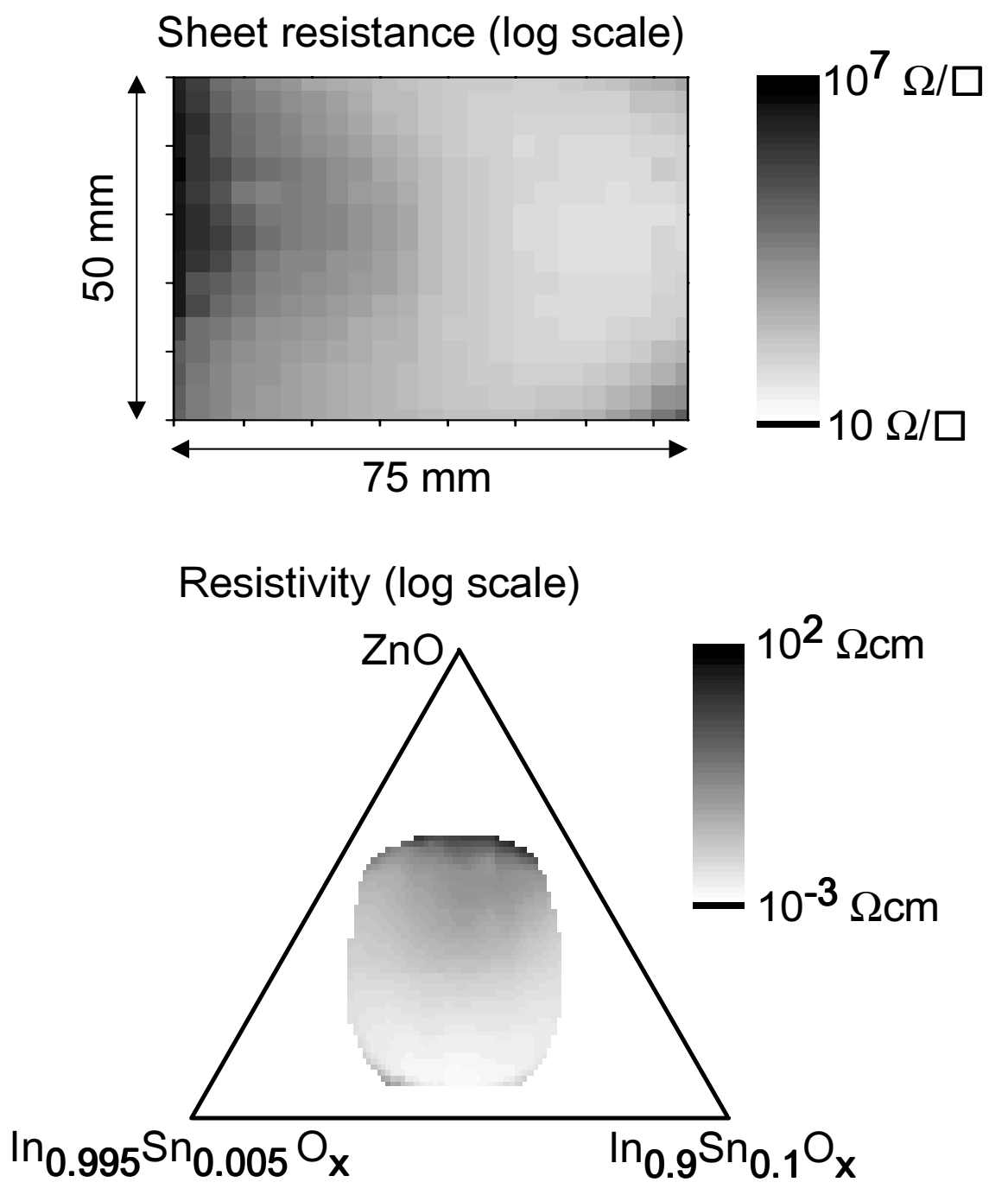

FIG. 7. Top: sheet resistance measured on a transparent conducting oxide thin film containing $SnO_2$, $In_2O_3$, and ZnO. Bottom: same data plotted in terms of resistivity on the pseudo-ternary phase diagram.

For three randomly chosen positions on this sample, indicated in Fig. 6b by squares, the composition was determined using RBS. The results thereof are shown as solid circles in the phase diagram. Also indicated (as crosses) are the results of the composition calculations based on the three parameters *n, r,* and *T*, for each material.

Clearly, the agreement between prediction and measurement is very satisfactory, especially when considering that the approach allows the user to "zoom in" on a subset of the phase diagram. This would be useful in particular for the study of "sharp" dependencies as a function of composition.

## IV. APPLICATION TO TRANSPARENT CONDUCTING OXIDES

Transparent conducting oxides have enjoyed significant interest in recent years, particularly for applications in flat-panel displays and for solar energy conversion devices. Growth methods that are applicable to low-temperature growth of these materials are particularly important, because the relevant substrates are often heat-sensitive.

PLD-growth of high-quality ITO has been reported by various groups,[24-26] and numerous studies have explored transparent oxides of similar compositions, such as $SnO_2$-ZnO,[27,28] $Ga_2O_3$-$In_2O_3$-$SnO_2$,[29] and $Ga_2O_3$-$In_2O_3$-ZnO.[30]

To demonstrate the principle of the PLD-CCS approach, the phase diagram consisting of $SnO_2$, $In_2O_3$, and ZnO, was chosen. However, for this particular system, the compositions with relatively small amounts of $SnO_2$ are of most interest, and it is therefore useful to be able to "zoom in" to that range.

To select the range of Sn:In ratios of interest, two ITO targets, namely $In_{0.995}Sn_{0.005}O_x$ and $In_{0.9}Sn_{0.1}O_x$ were chosen rather than the "true" end member $In_2O_3$ and $SnO_2$. The mixed film was again deposited onto a 2" by 3" glass slide, and the sheet resistance was determined by a standard four-probe technique at 368 evenly spaced points. The results are shown in Fig. 7. The observed variations in sheet resistance are due largely to the varying composition, but also contain a contribution from the non-uniform thickness (similar to the case of Fig. 6b). However, knowing the film thickness at each point (again by calculating the deposition rate profiles based on the parameters *n, r,* and *T* for each constituent), the resistivity can be calculated for each composition, hence eliminating the thickness dependence in the raw data. The resistivity can then be plotted as a function of composition, as is shown in the lower part of Fig. 7. Data of this type combined with optical transmission data (which can easily be obtained by a scanning method across a sample) will provide a useful tool for the discovery of new transparent conducting materials.

## V. FURTHER WORK

As a consequence of the fact that no post-annealing process is required, the approach can easily be applied to the growth of superlattices, where either the composition within each layer can vary across the sample, or where each layer is deposited with a non-uniform thickness, leading to a stacking with varying periodicity or varying ratio between the individual layer thicknesses. Initial results are published elsewhere.[31]

Further work is currently in progress to develop an algorithm to determine the composition without the need of reference samples. As first proposed by Hanak,[9] this can be done by thickness measurements at the perimeter of the actual sample and a least-square fitting procedure to yield the parameters that completely describe the composition dependence.

## ACKNOWLEDGMENTS

The authors would like to acknowledge P.M. Piccoli of the University of Maryland for the EDX measurements, and C. Woody White and David B. Poker of the Oak Ridge National Laboratory for the RBS analysis.